\documentclass[prb,twocolumn,showpacs,preprintnumbers,amsmath,amssymb]{revtex4}
\usepackage{amsfonts}
\usepackage{bbm}
\usepackage{mathrsfs}
\usepackage{amsmath}
\usepackage[dvips]{graphics,color}
\usepackage{graphicx}
\usepackage{epstopdf}
\usepackage{dcolumn}
\usepackage{bm}
\usepackage{longtable}
\usepackage{graphics}
\usepackage{amssymb}
\usepackage{xspace}
\usepackage{epsfig}
\usepackage{subfigure}
\usepackage{dcolumn}
\usepackage{multirow}
\makeatletter

\newcommand{\Rmnum}[1]{\expandafter\@slowromancap\romannumeral #1@}
\makeatother

\begin{document}
\title{Hund's Rules for the $N=0$ Landau Levels of Trilayer Graphene}
\author{Fan Zhang$^{1,2}$}
\email{zhf@sas.upenn.edu}
\author{Dagim Tilahun$^{2,3}$}
\author{A. H. MacDonald$^{2}$}
\affiliation{
$1$ Department of Physics and Astronomy, University of Pennsylvania, Philadelphia, PA 19104\\
$2$ Department of Physics, University of Texas at Austin, Austin TX 78712\\
$3$ Department of Physics, Texas State University, San Marcos, TX 78666}
\date{\today}
\begin{abstract}
The $N=0$ Landau levels of ABC and ABA trilayer graphene both have approximate $12$ fold
degeneracies that are lifted by interactions to produce strong quantum Hall effects (QHE) at all
integer filling factors between $\nu=-6$ and $\nu=6$.
We discuss similarities and differences between the strong-magnetic-field weak-disorder
physics of the two trilayer cases, and between trilayer and bilayer cases.
These differences can be used to identify the stacking order of high-quality trilayer samples
by studying their quantum Hall effects.
\end{abstract}
\maketitle
\section{Introduction}
Recent experimental advances\cite{Review_graphene} have realized a new family of quasi
two-dimensional electron systems (2DESs) consisting of two or more graphene layers.
Many of the unusual properties of these systems are related to momentum-space textures in the low-energy
quasiparticle bands,
and to the correspondingly enhanced $N=0$ (zero energy) Landau level (LL) degeneracies
that appear when an external magnetic field is applied.
As in the case of semiconductor 2DESs, Coulomb interactions break LL degeneracies
when disorder is weak\cite{Coulomb,spin2} and filling factors are close to integer values.
This phenomenon is sometimes referred to as quantum Hall ferromagnetism because its simplest\cite{pseudospin}
manifestation is spontaneous spin-polarization in the conduction bands of zincblende semiconductors at odd integer filling factors.
In few-layer graphene 2DESs,
LL degeneracies are larger because of the presence
of two low-energy valleys ($K$ and $K'$)\cite{QH_Novoselov,QH_Kim,QH_Kentaro},
and for $N=0$ also because of the momentum-space pseudospin textures.

In bilayer graphene, the additional degrees of freedom can
lead to canted antiferromagnetism\cite{SQH_Zhang,Kharitonov,LAF,preparation},
and valley and orbital order\cite{QHF_Yafis,QHF_Shizuya,QHF_Yacoby,QHF_Kim}
among other\cite{AB_Novoselov,AB_McCann,Dagim} phenomena.
The trilayer graphene\cite{ABC_Zhang,ABC_McCann,Mul_Peeters,ABA_McCann} system, which has become experimentally realized\cite{Bao,SiC_ABC,ABA,ABC_Bao,Trilayer_LYZhang,Trilayer_Kumar,Trilayer_Jhang,Trilayer_Lui,Trilayer_Eisenstein,Trilayer_Russo} only recently,
presents an even richer many-electron physics landscape. Both ABC (chiral) and ABA stacked trilayers are expected to have integer Hall plateaus at $\nu=\pm4(n+{3}/{2})$.
The $\nu=-6$ and $\nu=6$ plateaus are separated by a set of $12$ approximately degenerate $N=0$ LLs, the trilayer duodectet, due to a combination of the $4$-fold spin-valley degeneracy common to all levels, and
trilayer momentum space pseudospin textures\cite{ABC_Zhang,ABC_McCann}.
However, as we shall explain, the role of electron-electron interactions is quite different in the two cases because of differences\cite{ABC_Bao,SQH_Zhang,new2} in the spatial and orbital contents of the $N=0$ LL wavefunctions.

\begin{figure}[t]
\label{fig:one}
\centering \scalebox{0.4} {\includegraphics* {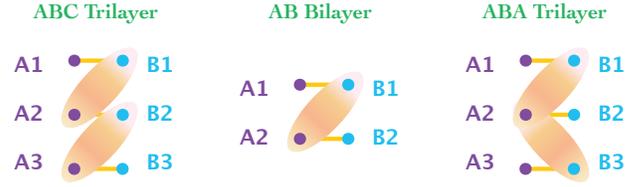}}\caption{\label{fig:structure} {(Color online) Schematic electronic structure of
few-layer graphene.  The sites coupled by the strongest interlayer bonds characterized by the $\gamma_1$ hopping
parameter are indicated by shading and have little weight in
$N=0$ Landau levels.}}
\end{figure}
ABC trilayers have $N=0$ layer wavefunctions
that are localized on the $A_1$ site in Fig.~\ref{fig:one} for
one valley and on the $B_3$ site for the other.
ABC trilayers are like AB bilayers
in that they have inversion symmetry protected\cite{RG_Zhang} degeneracy, but
they differ in that they have $N=0$ LLs with $n=2$ orbital character, in addition to $n=0$ and
$n=1$.  Their quantized LL energies are $E_{\rm n}^{(3)}\sim B^{3/2}\sqrt{n(n-1)(n-2)}$.
This property causes LL mixing effects to vanish strongly at very high magnetic fields.
The $N=0$ LLs of ABA trilayers, on the other hand, consist of $n=0,1$ orbitals on the layer symmetric states $|A_1\rangle+|A_3\rangle$ and $|B_2\rangle$, and only $n=0$ orbitals on the layer antisymmetric states $|A_1\rangle-|A_3\rangle$ and $|B_1\rangle-|B_3\rangle$.
The $N\ne 0$ LL spectrum can be viewed as a superposition
of monolayer-like LLs with $E_{\rm n}^{(1)}\sim B^{1/2}\sqrt{n}$ and bilayer-like LLs
with $E_{\rm n}^{(2)}\sim B\sqrt{n(n-1)}$.  This structure leads to LL crossing\cite{ABA_McCann,ABA,Trilayer_Yuan}
and for $N\ne 0$ increases the quantitative importance of LL mixing at stronger fields compared to the ABC case.

Our description of the $N=0$ physics of ABC and ABA trilayer graphene
starts from a model that includes only nearest neighbor intra-layer and inter-layer hopping and focuses on energies
smaller than the inter-layer hopping strength $\gamma_1$.  By treating other hoppings perturbatively at
low-energies, it is possible to derive effective Hamiltonians for $N=0$ LLs that are valid at realistic magnetic
field strengths. In Section II of this article we discuss $N=0$ quantum Hall ferromagnetism in
ABC trilayer graphene, predicting a simple set of Hund's rules that apply in the absence of
inversion-symmetry breaking external electric fields.  We then address the phase transitions
that can be driven experimentally by applying electric fields across the trilayers using top and bottom gates.
The interaction physics of the $N=0$ level in ABA graphene, which is discussed in
Section III, is more complex even in the absence of an external electric field.  We conclude
in Section IV with a discussion of prospects for interesting experimental discoveries in these
systems.

\section{ABC-stacked trilayer graphene}
The low-energy model for ABC trilayers\cite{ABC_Zhang,ABC_McCann} in
the $A_1$-$B_3$ representation is:
\begin{eqnarray}
\mathcal{H}_{\rm ABC}&=&
\frac{v_0^3}{\gamma_1^2}\left( \begin{array}{cc}
0 & \pi^{\dag\,3}\\
 \pi^3 & 0\\
\end{array} \right)\nonumber\\
\nonumber\\
&+&u_{\rm d}\left[
\left( \begin{array}{cc}
1 & 0\\
 0 & -1\\
\end{array} \right)
-\frac{v_0^2}{\gamma_1^2}\left( \begin{array}{cc}
\pi^{\dag}\pi & 0\\
 0 & -\pi\pi^{\dag}\\
\end{array} \right)\right]\,.
\label{eq:abc}
\end{eqnarray}
In Eq.(\ref{eq:abc})  ${\bm \pi}={\hbar\bm k}+e{\bm A}/c$ is the 2D kinetic momentum, $\pi=\tau^{\rm z}\pi_{\rm x}+i\pi_{\rm y}$
where $\tau^{\rm z}=\pm 1$ represents valley $K$ and $K'$,
 $v_0$ is the monolayer Dirac velocity, $\gamma_1\sim 0.5$ eV\cite{ABC_Zhang,Trilayer_LYZhang},
 and $2u_{\rm d}$ is the site-energy difference between $A_1$ and $B_3$.
This model is approximately valid for $\hbar v_0/\ell_{\rm B} \ll \gamma_1$ where $\ell_{\rm B}=25.6\,{\rm nm}/\sqrt{B[T]}$ is the magnetic length, {\em i.e.} for fields much smaller than $\sim 400$ T.
 For $u_{\rm d}=0$, $\mathcal{H}_{\rm ABC}$ has three zero energy eigenstates for
each value of $\tau_{z}$.  The corresponding eigenstates, which play the key role in ABC trilayer
quantum Hall ferromagnetism have the form $(0, \phi_{\rm n})$ for valley $K$ and $(\phi_{\rm n},0)$ for valley $K'$.
Here $n=0,1,2$ is an orbital label and $\phi_{\rm n}$ is the $n$th LL eigenstate of an
ordinary 2DEG.  The ABC trilayer duodectet is the direct product of this LL orbital triplet, a
real spin doublet, and a {\em which-layer} (or equivalently which-valley) doublet.

Gaps can open within the duodectet, giving rise to additional integer quantum Hall effects, due to single-particle terms that
lift degeneracies or due to interactions that break symmetries.  For example a Zeeman term
splits spin levels by $2E_{\rm ZM}=g\mu_{\rm B}B=0.116\times B[T]$ meV,
while $u_{\rm d}$ splits both layers and orbitals: $E_{\rm LL}^{n}=-\tau^{\rm z}u_{\rm d}(1-n(\hbar\omega_{3}/\gamma_1)^{{2}/{3}})$.
The notation used in this equation is motivated by viewing the ABC multilayer graphene Hamiltonian as the $J=3$ generalization of Hamiltonians that apply to single-layer graphene for $J=1$ and to bilayer graphene for $J=2$. In each case
$\hbar\omega_{\rm J}$ is proportional to the energy of the $N=1$ LL.
$\hbar\omega_{\rm J}=(\sqrt{2}v_0\hbar/\gamma_1 \ell_{\rm B})^{\rm J}\gamma_1$
is the cyclotron frequency of quasiparticles in chirality $J$ 2DES's.     The Zeeman energy is invariably small compared to
the interaction scale $e^2/\epsilon \ell_{\rm B}$, whereas $u_{\rm d}$ can be large enough to completely reorganize
the duodectet.

The full electron-electron interaction Hamiltonian in the duodectet subspace is
readily available given the simple form factors of the $n=0,1,2$ Landau levels.
In our discussion of $N=0$ quantum Hall ferromagnetism we will nevertheless use the
Hartree-Fock approximation which is expected to be accurate\cite{Schliemann}
at the integer filling factors of interest, and more physically transparent.
We solve the Hartree-Fock equations in the
duodectet subspace allowing any symmetry, other than translational invariance,
to be broken. The Hartree-Fock Hamiltonian is expressed in terms of the duodectet
density-matrix which must be determined self-consistently at each filling factor.
When we include only interactions\cite{LLprojectioncaveat} within the $N=0$ manifold
we find the quasiparticle Hamiltonian with Hartree and exchange interaction contributions is
\begin{align}
\langle\alpha n s|\mathcal{H}_{\rm ABC}^{\rm HF}|\beta n' s'\rangle=&\left(E_{\rm LL}^{n} \delta_{\rm ss'}-E_{\rm ZM}\sigma^{\rm z}_{\rm ss'}\right)\delta_{\rm nn'}\delta_{\rm \alpha\beta}\nonumber\\
&+E_{\rm H}(\Delta_{\rm B}-\Delta_{\rm T})\tau^{\rm z}_{\rm \alpha\beta}\delta_{\rm ss'}\delta_{\rm nn'}\nonumber\\
&-E_{\rm F}\sum_{\rm n_1 n_2}X^{\rm \alpha\beta}_{\rm n n_{2}, n_{1} n'}\Delta_{\rm \alpha n_2 s}^{\rm \beta n_1 s'}\delta_{\rm ss'}\,,
\end{align}
where we have used the labels $n=0,1,2$ for orbitals,
index $s$ for spin, and the Greek indices $\alpha,\beta=B(K)$ or $T(K')$ for layer(valley).
Here $E_{\rm F}={e^2}/{\varepsilon \ell_{\rm B}}$
characterizes the overall strength of the exchange interactions,
$E_{\rm H}=({2d}/{2\ell_{\rm B}})E_{\rm F}$ captures the capacitive electrostatic
energy associated with shifting charge between the top and bottom layers,
and $d=0.335$ nm is the separation between adjacent graphene layers.
The layer projected charge density is
\begin{align}
\Delta_{\rm \alpha}=\sum_{\rm n s}\Delta_{\rm \alpha n s}^{\rm \alpha n s}\,,
\end{align}
and the density matrix is
\begin{align}
\Delta_{\rm \alpha n s}^{\rm \beta n' s'}=\langle c^{\dag}_{\rm \beta n' s'}c_{\rm \alpha n s}\rangle\,.
\end{align}
The dimensionless exchange integrals $X^{\rm \alpha\beta}_{\rm n n_{2}, n_{1} n'}$
reflect the form factors associated with $n=0,1,2$ orbitals.
We find that
\begin{align}
\label{eq:X}
X^{\rm \alpha\beta}_{\rm n, n_{2}, n_{1} n'}=\int\frac{d^2 {\bm k}}{(2\pi)^2}F_{\rm n n_2}(-{\bm k})F_{\rm n_1 n'}({\bm k})\frac{2\pi e^2}{\varepsilon k}\frac{\eta^{\rm \alpha\beta}}{E_{\rm F}}\,,
\end{align}
where $\eta^{\rm \alpha\beta}$ becomes $1$ for $\alpha=\beta$ and $e^{\rm -2kd}$ for $\alpha\neq\beta$.
Here $F_{\rm nn'}({\bm k})$ are the ordinary\cite{2DEGHF} 2DEG LL form factors
which can be expressed in terms of associated Laguerre polynomials that capture the spatial profile of the LL wavefunctions.
The exchange integrals are non-zero only for $n-n'=n_2-n_1$; numerical values of the
relevant form factors for the case $d=0$ are listed in Table~\ref{table:one}.

\begin{table*}[t!]
\caption{Summary of the numerical values of exchange integrals.
The intralayer integrals are independent of magnetic fields while the interlayer ones
decline slowly as the field strength increases.  Numerical values are reported for $B=20$ T and $B=30$ T.
We have defined indices $a=|n-n_2|$, $b=\min(n_1,n')$ and $c=\min(n,n_2)$.
Exchange integrals are symmetric between $b$ and $c$.
For the interlayer exchange integrals the values reported here are for neighboring layer separation $d=0.335$ nm.}
\newcommand\T{\rule{0pt}{3.1ex}}
\newcommand\B{\rule[-1.7ex]{0pt}{0pt}}
\begin{scriptsize}
\centering
\begin{tabular}{c c c | c | c | c || c c c | c | c |c || c c c | c | c | c}
\hline\hline\T
\multirow{2}{*}{$a$}\; & \multirow{2}{*}{$b$}\; & \multirow{2}{*}{$c$}\; & Intralayer & Interlayer & Interlayer & \multirow{2}{*}{$a$}\; & \multirow{2}{*}{$b$}\; & \multirow{2}{*}{$c$}\; & Intralayer & Interlayer & Interlayer &
\multirow{2}{*}{$a$}\; & \multirow{2}{*}{$b$}\; & \multirow{2}{*}{$c$}\; & Intralayer & Interlayer & Interlayer\\
 & & & Any $B$ & $B=20$ T & $B=30$ T & & & & Any $B$ & $B=20$ T & $B=30$ T & & & & Any $B$ & $B=20$ T & $B=30$ T\T\\[3pt]
\hline
 0 & 0 & 0 & 1.2533 & 1.1444 & 1.1219 & 1 & 0 & 0 & 0.6267 & 0.5215 & 0.5008 & 2 & 0 & 0 & 0.4700 & 0.3676 & 0.3481\T\\
 0 & 0 & 1 & 0.6267 & 0.6229 & 0.6211 & 1 & 0 & 1 & 0.2216 & 0.2177 & 0.2160 & 2 & 0 & 1 & 0.1357 & 0.1319 & 0.1302\T\\
 0 & 0 & 2 & 0.4700 & 0.4689 & 0.4684 & 1 & 0 & 2 & 0.1357 & 0.1348 & 0.1343 & 2 & 0 & 2 & 0.0720 & 0.0711 & 0.0708\T\\[5pt]
 0 & 1 & 0 & 0.6267 & 0.6229 & 0.6211 & 1 & 1 & 0 & 0.2216 & 0.2177 & 0.2160 & 2 & 1 & 0 & 0.1357 & 0.1319 & 0.1302\T\\
 0 & 1 & 1 & 0.9400 & 0.8365 & 0.8165 & 1 & 1 & 1 & 0.5483 & 0.4470 & 0.4280 & 2 & 1 & 1 & 0.4308 & 0.3315 & 0.3133\T\\
 0 & 1 & 2 & 0.5483 & 0.5434 & 0.5412 & 1 & 1 & 2 & 0.2398 & 0.2348 & 0.2325 & 2 & 1 & 2 & 0.1592 & 0.1542 & 0.1520\T\\[5pt]
 0 & 2 & 0 & 0.4700 & 0.4689 & 0.4684 & 1 & 2 & 0 & 0.1357 & 0.1348 & 0.1343 & 2 & 2 & 0 & 0.0720 & 0.0711 & 0.0708\T\\
 0 & 2 & 1 & 0.5483 & 0.5434 & 0.5412 & 1 & 2 & 1 & 0.2398 & 0.2348 & 0.2325 & 2 & 2 & 1 & 0.1592 & 0.1542 & 0.1520\T\\
 0 & 2 & 2 & 0.8029 & 0.7029 & 0.6844 & 1 & 2 & 2 & 0.4994 & 0.4010 & 0.3832 & 2 & 2 & 2 & 0.4027 & 0.3058 & 0.2887\T\\
\hline\hline
\end{tabular}
\end{scriptsize}
\label{table:one}
\end{table*}

\begin{figure}[t]
\centering \scalebox{0.62} {\includegraphics*[1.5in,3.30in][6.80in,8.00in]{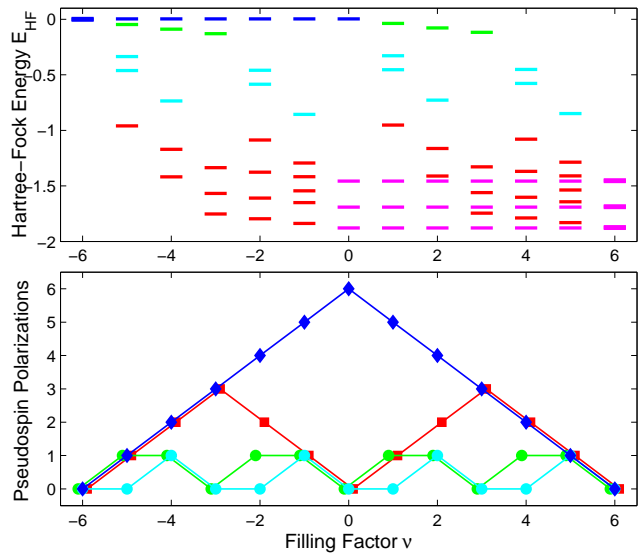}}\caption{\label{fig:abc} {(Color online) Upper panel:
filling factor dependence of the Hartree-Fock energies of occupied and unoccupied LLs for a balanced ABC trilayer at $20$ T.  Energies are in units of $(\pi/2)^{1/2}e^2/\varepsilon \ell_{\rm B}$.  Unoccupied LLs are sextets (blue), triplets (green), or singlets (cyan); occupied LLs are doublets (magenta) or singlets (red); LLs at $\nu=\pm 6$ are split only by the small Zeeman splitting.
These energies include only exchange interactions within the $N=0$ level.  Particle-hole symmetry within the $N=0$ space
is recovered if exchange interactions with remote Landau levels are included; these interactions do not influence the energy spacings
between Landau levels.
Lower panel: filling factor dependence of the polarizations of up spin (blue) and layer (red), and
population differences between $n=2$ and
$n=0$ (green) and $n=1$ (cyan) LL orbitals.}}
\end{figure}

The self-consistent solutions of the $N=0$ ABC Hartree-Fock equations for balanced ($u_{d}=0$)
ABC trilayers are summarized in Fig.\ref{fig:abc}
for a typical field of $B=20$ T.  Gaps appear at all intermediate integer
filling factors, and as expected are always much larger than $E_{\rm ZM}$.
The electronic ground state at duodectet filling factors from $\nu=-6$ to $6$
follows a Hund's rule behavior progression in which
spin polarization is maximized first, then layer (valley) polarization to the greatest extent possible,
and finally LL orbital polarization to the extent allowed by the first two rules.
In these calculations, the quasiparticle energies of spins along the field direction (say spin $\uparrow$)
are much lower than those for spin $\downarrow$ states because the spin-splitting is exchange enhanced.
Since $d/\ell_{\rm B}$ is always small in trilayer graphene systems, the exchange integrals that
favor spin-polarization are only slightly larger than those that favor layer polarization.
Because these differences and Zeeman energies are both small,
inter-valley exchange\cite{Jung} (which is neglected in the continuum model we employ here)
can potentialy alter the competition between spin and layer polarization compared to
the results reported in Fig.\ref{fig:abc}.  We return to this
point in the discussion section. For balanced ABC trilayers, layer-order appears as spontaneous inter-layer phase
coherence\cite{layer,interlayer,Exciton_Condensate} rather than layer polarization,
and to Hartree-Fock quasiparticles that have symmetric ($S$) or anti-symmetric ($AS$) layer states.
The three LL orbitals are filled in the order of increasing $n$ because
exchange integrals decrease with $n$ as indicated in Table~\ref{table:one}.
It follows that the first six filled LLs are $|S,0\uparrow\rangle$, $|S,1\uparrow\rangle$ and $|S,2\uparrow\rangle$ followed by the AS counterparts; the next six filled LLs are the spin $\downarrow$ states in the same order.
Here $S$ and $AS$ are orthogonal coherent bilayer states, chosen as symmetric ($S$) and
antisymmetric ($AS$) states as a matter of convenience.

These Hund's rules imply that the interaction driven
integer quantum Hall states are spin and pseudospin polarized at all the
intermediate integer filling factors from $\nu=-5$ to $5$, as summarized in Fig.\ref{fig:abc}.
All $11$ states are spin $\uparrow$ polarized ferromagnets with a maximum
spin polarization at $\nu=0$.
The LL orbital dependence of the microscopic Hamiltonian leads to orbital
polarization except at filling factors that are multiples of $3$.
In the absence of an interlayer potential which breaks inversion symmetry,
the layer pseudospin symmetry is broken by
states with spontaneous interlayer coherence which
open a gap at no cost in Hartree energy.
The Hartree energy and the difference between intralayer and interlayer exchange are both
the same order $\sim {d}/{\ell_{\rm B}}\cdot{e^2}/{\ell_{\rm B}}$, but
it turns out that Hartree energy slightly dominates over exchange difference favoring
spontaneous coherence state over competing layer polarized states.

We have so far neglected the strong dependence of
duodectet quantum Hall ferromagnets on an interlayer electric field.
A very small potential difference between the top and bottom layers is sufficient to
destabilize spontaneous-coherence states in favor of layer polarized states.
Quantum phase transitions {\em vs.}  $u_{d}$ are expected at all filling factors.
For $\nu=0$, for example,
there is a first-order phase transition from the spin polarized ferromagnetic state
to a layer(valley) polarized state when the potential difference between the top and bottom layers
becomes dominant over the exchange induced spin splitting. For $B=20$ T ($\varepsilon=1$)
we find that the layer polarized state has the lowest energy if the potential difference
between the top and bottom layers exceeds $0.2$ eV.
The gap between the highest occupied LL $|T,2,\uparrow\rangle$
and the lowest unoccupied LL $|B,0,\downarrow\rangle$ is $0.46$ eV for $u_{d}=0$ and is reduced to
$0.28$ eV at the phase transition.
We find that the critical screened electric field for this first order phase transition
 is $\sim 15$ mV/(nm$\cdot$T) for $\varepsilon=1$, comparable to the experimental value
 of the corresponding transition field in
 bilayers.\cite{cdw1,cdw2,cdw3,cdw4}
 The linear dependence of the critical field on $B$ follows from the fact that
 all relevant microscopic energy sales (the bias supported
 energies $E_{\rm H}$ and $E_{\rm LL}$, the difference between intralayer and interlayer exchanges, and the
 Zeeman energy $E_{\rm ZM}$) are all linear functions of $B$.

In high mobility and low disorder suspended samples, we anticipate that the $\nu=\pm 6,0$ states and the bias supported $\nu=\pm 3$ states, persist down to nearly zero magnetic field exhibiting spontaneous quantum Hall effects.\cite{SQH_Zhang,ABC_Bao}
ABC trilayers are highly susceptible to broken symmetries that are accompanied by large momentum space Berry curvatures and different types of topological order.\cite{SQH_Zhang}  Remote weaker hopping processes\cite{ABC_Zhang} in ABC trilayers that we have
ignored so far could influence the field to which the quantum Hall states survive.
The most important of these appears likely to be the trigonal warping energy scale$\sim7$ meV.\cite{ABC_Zhang}
The study of weak field quantum Hall effects in ABC trilayers should shed light on the
nature of the $B=0$ ground state.\cite{SQH_Zhang}

\section{ABA-stacked trilayer graphene}
Unlike ABC trilayer quasiparticles with pure chirality, the band structure of an ABA trilayer consists of a massless monolayer ($J=1$) and a massive bilayer ($J=2$) subbands.  Unbiased ABA trilayers have mirror symmetry with respect to the middle layer and their low energy physics is governed by\cite{ABA_McCann}
\begin{eqnarray}
\mathcal{H}_{\rm ABA}&=&
\left[v_0\left( \begin{array}{cc}
0 & \pi^{\dag}\\
 \pi & 0\\
\end{array} \right)
+\left( \begin{array}{cc}
-\frac{\gamma_2}{2} & 0\\
 0 & \delta'-\frac{\gamma_5}{2}\\
\end{array} \right)\right]_{\rm J=1}\nonumber\\
\nonumber\\
&\oplus&
\left[\frac{-v_0^2}{\sqrt{2}\gamma_1}\left( \begin{array}{cc}
0 & \pi^{\dag\,2}\\
 \pi^2 & 0\\
\end{array} \right)
+\left( \begin{array}{cc}
\frac{\gamma_2}{2} & 0\\
 0 & 0\\
\end{array} \right)\right]_{\rm J=2}\,,
\label{eq:aba}
\end{eqnarray}
where the $J=1$ subbands are layer antisymmetric states $|A_1\rangle-|A_3\rangle$ and $|B_1\rangle-|B_3\rangle$ while the $J=2$ subbands are layer symmetric states $|A_1\rangle+|A_3\rangle$ and $|B_2\rangle$.  Note that the
effective interlayer hopping energy is enhanced by a factor of $\sqrt{2}$ compared to the AB bilayer
case.  Again, the mirror symmetry leads to spontaneous coherence states.  The next-nearest-neighbor
tunneling processes play a more essential role than in the
ABC trilayer case.  Here
$\gamma_2=-20$ meV, $\gamma_5=40$ meV and $\delta'=50$ meV\cite{ABA_McCann,ABA} lead to band gaps for the
$J=1$ and $J=2$ chiral branches separately, but no direct gap overall.  When Zeeman splittings and
interactions ignored, the duodectet has four-fold degenerate $J=2$ eigenenergies at $\epsilon \simeq -10,0 {\rm meV}$ and
two-fold degenerate $J=1$ eigenenergies at $\epsilon \simeq 10,30 {\rm meV}$.
These energies are equal to the $B=0$ band energy extrema and are independent of magnetic field.
The single-particle splitting of
$N=0$ can therefore lead to quantum Hall effects at $\nu=-2$, $\nu=2$ and $\nu=4$.

The self-consistent Hartree-Fock Hamiltonian that describes the broken symmetry ABA trilayer $N=0$ duodectet is
\begin{align}
\label{eq:abaB}
\langle i\, n_{\rm i}\, s|\mathcal{H}_{\rm ABA}^{\rm HF}|j\, n_{\rm j}\, s'\rangle\!=\!&\left(E_{\rm LL}'\delta_{\rm ss'}-E_{\rm ZM}\sigma^{\rm z}_{\rm ss'}\right)\delta_{\rm n_i n_j}\delta_{\rm ij}\nonumber\\
&+\!\frac{E_{\rm H}}{2}\Delta_{\rm B_2}\left(2\delta_{\rm B_2,i}-1\right)\delta_{\rm ss'}\delta_{\rm n_i n_j}\delta_{\rm ij}\nonumber\\
&-\!E_{\rm F}\sum_{\rm n_1 n_2}X^{\rm i j}_{\rm n_i n_{2}, n_{1} n_j}\Delta_{\rm i\, n_2 s}^{\rm j\, n_1 s'}\delta_{\rm ss'}\,,
\end{align}
where the LL index $n_{\rm i}$ range depends on the atomic orbital $i$, {\em i.e.}, $n_i=0$ only for $J=1$ branch while $n_i=0,1$
for $J=2$ branch.  The exchange integrals $X^{\rm i j}$ are
still defined by Eq.(\ref{eq:X}) but with a more general definition $\eta^{\rm ij}=\sum_{\rm m}c_{\rm m}V_{\rm m}/V_0$,
where $V_{0,1,2}$ respectively denote intralayer, nearest interlayer, and next-nearest interlayer Coulomb interactions,
and $c_{\rm m}$ is obtained from the interaction matrix element
decomposition $\langle i j|V|i j \rangle=\sum_{\rm m}c_{\rm m}V_{\rm m}$.
Otherwise the notation in Eq.(\ref{eq:abaB}) is the same as in the ABC case.
Since it is already quite complicated,
we focus on the balanced case\cite{new1,new3} with $u_{\rm d}=0$.
$E_{\rm LL}$ is therefore absent and replaced by $E_{\rm LL}'$ which is
given by the $\vec{\pi}=0$ eigenvalues of Eq.(\ref{eq:aba}).

\begin{figure}[t]
\centering \scalebox{0.53}
{\includegraphics*[0.95in,2.75in][7.25in,8.30in]{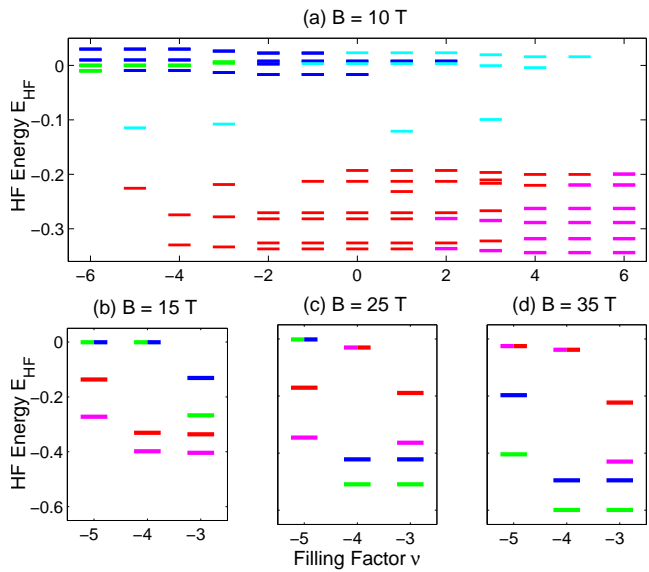}}
\caption{\label{fig:aba} {(Color online) (a) Filling factor dependence of the Hartree-Fock (HF) energies (eV) of occupied and unoccupied LLs for a balanced ABA trilayer at $10$ T.  Unoccupied LLs are quartets (green), doublets (blue), or singlets (cyan); occupied LLs are doublets (magenta) or singlets (red). (b)-(d) Field dependence of the HF energies (eV) of the lowest four LLs at $\nu=-5,-4$ and $-3$.  Results for magnetic field $B$ equal to $15$ T are shown in (b), for $25$ T in (c) and for $35$ T in (d).  Magenta and red denote the $n=0$ and $1$ orbitals respectively of LL $|A_1+A_3\,\uparrow\rangle$; green and blue denote the $n=0$ and $1$ orbitals respectively of LL $|B_2\,\uparrow\rangle$.
The Zeeman splitting is small enough to ignore in this figure and we assumed that $\varepsilon=1$.}}
\end{figure}

We find that the ABA duodectet intermediate filling factor states
follow the Hunds rule for maximum spin-polarization, but are otherwise strongly
influenced by the single-particle gaps.
The $J=2$ LLs, which have lower single-particle energies are filled first.
When the magnetic field is smaller than $B_{\rm c1}=17$ T ($\varepsilon=1$ hereafter),
the first six filled LLs are all majority spin levels:
$|A_1+A_3\,\uparrow\rangle$ with $n=0$ and $1$, then $|B_2\,\uparrow\rangle$ with $n=0$ and $1$, and lastly $|A_1-A_3\,\uparrow\rangle$ and $|B_1-B_3\,\uparrow\rangle$ with $n=0$.
The next six filled LLs are the spin $\downarrow$ states in the same order.
Within a given spin the LL filling pattern follows the order suggested by the
single-particle physics.
The Hartree-Fock theory predictions for field strength $B=10$ Tesla are summarized in Fig.\ref{fig:aba}(a).
At $\nu=-5$ and $\nu=-4$ $|A_1+A_3\,\uparrow\rangle$ LLs are occupied first.
In addition to being favored by the single-particle energies, this choice
gains exchange energy and minimizes Hartree energy.
At $\nu=-3$ the $|B_2\,0,\uparrow\rangle$ LL is then occupied in agreement with the
single-particle Hamiltonian, but also lowering the interaction energy.
Notice the the gaps between occupied and empty states, which are
important for transport, are greatly enhanced relative
to the single-particle gaps at all intermediate filling factors.  The smallest Fermi level gaps occur
at filling factors $\nu=-5$ and $\nu=-3$, for which occupied and empty states are separated
by a gap between $n=0$ and $n=1$ orbitals with the same layer structure.
These two filling factors should therefore have the weakest quantum Hall effects in
the weak-disorder samples to which the present calculations apply.

Although the LL filling order is universal for $\nu>-3$,
there is some field dependence for $\nu=-5,-4$ and $-3$ as indicated in Fig.\ref{fig:aba}.
When the field strength exceeds $B_{\rm c1}=17$ T a LL reordering occurs at $\nu=-4$.
We find that the pair of LLs $|B_2\,\uparrow\rangle$ are then filled before the $|A_1+A_3\,\uparrow\rangle$
pair.  This transition occurs when Coulomb interaction physics overwhelms
the single-particle next-nearest layer tunneling effect.
More localized layer orbitals are preferred when interactions dominate
because intralayer exchange energy gains dominate interlayer exchange and
Hartree energy costs.
When the field is further strengthened to $B_{\rm c2}=33$ T LL
$|B_2\,0\,\uparrow\rangle$ is filled before $|A_1+A_3\,0\,\uparrow\rangle$ , at $\nu=-5$.
The two critical fields are roughly determined by the
condition $E_{\rm H}/2\sim\Delta E_{\rm LL}'$ which coincides
approximately with the  self-consistent
numerical results for $B_{\rm c1}$ and $B_{\rm c2}$.

We have focused here on interaction physics in the $N=0$ LL's of ABC and ABA trilayer graphene.
Interesting effects are nevertheless particularly likely for larger $N$ in the ABA case.
Unlike ABC trilayers where only $J=3$ bands are present at low energies, both $J=1$ and $2$ subbands appear at all energies in ABA trilayers, leading to LL crossing\cite{ABA_McCann,ABA,Trilayer_Yuan} between the two chiral branches when $\omega_1/\omega_2=\sqrt{n_2(n_2-1)/2n_1}$ is satisfied.  We anticipate
that Coulomb physics will open gaps at the LL crossing points, whose characteristics would depend on the orbital indices of the crossing LLs.  Because of the LL crossing effects present even without interactions,
the sequence of the plateaus in ABA trilayers highly depend on the magnetic field strength.
\newline\indent

\section{Discussion}

Single-layer, bilayer, and ABC and ABA trilayer graphene are a family of related but distinct
two-dimensional electron systems.  Single-layer and bilayer graphene have four and eight fold
Landau level degeneracies that are lifted by interactions in the quantum Hall regime when disorder
is sufficiently weak, giving rise to a rich variety of broken symmetry states.  In this paper we
have explored the way in which interactions alter the quantum Hall effect in both ABC and
ABA trilayer graphene in the range of filling factors between $\nu=-6$ and $\nu=6$, over which
their twelve-fold $N=0$ Landau level is filled.  As in the single-layer and bilayer cases our
calculations suggest that quantum Hall effects will eventually occur at all intermediate filling
factors as sample quality improves to lessen the role of disorder.
There are a number of clear differences
between the ABC and ABA cases which could provide a practical alternative to Raman scattering for
identifying the stacking orders of high quality trilayer samples: i)  Because the $N=0$ layer orbitals are
localized either in the top or bottom layer in the ABC case, while they are often spread across two or
more layers in the ABA case, interaction effects are stronger for ABC trilayers and more likely to
induce intermediate filling factor quantum Hall effects. ii) Perpendicular electric fields have
relatively simple consequences in ABC trilayers, opening up gaps between layers that will
strengthen the quantum Hall effect at $\nu=\pm 3$ and $\nu=0$, but have a much more complex
behavior with many quantum phase transitions in the ABC case.
For ABC trilayers there is a single sharp phase transition from a spin-polarized (at high field) or antiferromagnetic (at low field) $\nu=0$ state to
a layer polarized state with a critical field strength that is linear in B.
iii) In ABA samples on substrates, the single-particle splitting of
$N=0$ LLs might lead to quantum Hall effects at $\nu=-2$, $\nu=2$ and $\nu=4$ with gaps $\sim 10-20$ meV before the interaction driven quantum Hall states appear at larger fields.
iv) There are phase transitions in ABA trilayers at  $\nu=-5, -4, -3$ between low-field states in which
single-particle physics determines the order in which LL's are occupied, to high-field states
in which more localized orbitals with larger exchange energies are occupied first.
v) In high quality suspended ABC trilayers, $\nu=0, \pm 3, \pm 6$ states persist down to zero magnetic field and smoothly connect with the electron-electron interaction driven broken symmetry states\cite{SQH_Zhang}.
We expect that these differences will become apparent in future experiments.

In this paper we have reported on Hartree-Fock ground states and quasiparticle energies as a function of filling factor
predicted by calculations which were restricted only by the assumption of translational invariance but otherwise
allowed any symmetries of the Hamiltonian to be broken.
We find that in the strong-magnetic-field weak-disorder case the LL filling order tends to follow a pattern in which states of a particular spin are filled first, and within a spin states with different valley content next.  For ABA trilayers, however, single-particle physics which places a bilayer-like group of states lower in energy than a monolayer-like group of states, can play an important role.  The rule which favors spin-polarization over valley (or equivalently layer) polarization is weak because the layers in this system are quite closely spaced, and it is possible that this rule will be overturned by lattice effects - as it appears to be in bilayers\cite{Jung}.
When $M$ of the $12$ $N=0$ Landau levels are
occupied, there will be a set of $M \times (12-M)$ collective modes whose energies approach
differences between quasiparticle energies in the limit of long-wavelength.  Many of these collective modes,
which can be studied\cite{Yafis_ABC} using time-dependent versions of the calculations reported in this paper,
will couple to long-wavelength electromagnetic probes which would provide information about the
many-body ground state that is complementary to the transport activation gaps most related to the present study.

\acknowledgements
F.Z. gratefully acknowledges helpful discussion with C.N. Lau, I. Zaliznyak, J. Velasco Jr, W. Bao and K. Zou.
This work has been supported by Welch Foundation grant TBF1473, NRI-SWAN, DOE grant Division of
Materials Sciences and Engineering DE-FG03-02ER45958, and ARO W911NF-09-1-0527.


\begin{thebibliography}{100}
\bibitem{LLprojectioncaveat}  The neglected exchange interactions with remote
 Landau levels shift energies by a constant which does not influence physical
 properties.  In the projected calculation quasiparticle-energies are
 not particle-hole symmetric as the $N=0$ level is filled up, but
 energy-differences are.
\end{thebibliography}
\end{document}